\documentclass[preprint,pra,aps,nofootinbib,longbibliography]{revtex4-2}
\usepackage{graphicx}
\usepackage{tabularx}
\usepackage{longtable}
\usepackage{booktabs}
\usepackage[table]{xcolor}
\usepackage{amssymb}
\usepackage{bm}

\newcommand{\Lim}[1]{\raisebox{0.5ex}{\scalebox{1.0}{$\displaystyle \lim_{#1}\;$}}}

\begin{document}

\title{Gaussian Basis Functions for a Polymer Self-Consistent Field Theory of Atoms}
\author{Phil A. LeMaitre}
\affiliation{Department of Physics \& Astronomy, University of Waterloo, 200 University Avenue West, Waterloo, Ontario, Canada N2L 3G1}
\author{Russell B. Thompson}
\email{thompson@uwaterloo.ca}
\affiliation{Department of Physics \& Astronomy and Waterloo Institute for Nanotechnology, University of Waterloo, 200 University Avenue West, Waterloo, Ontario, Canada N2L 3G1}
\date{\today}
\begin{abstract}
A representation of polymer self-consistent field theory equivalent to quantum density functional theory is given in terms of non-orthogonal basis sets. Molecular integrals and self-consistent equations for spherically symmetric systems using Gaussian basis functions are given, and the binding energies and radial electron densities of neutral atoms hydrogen through krypton are calculated. An exact electron self-interaction correction is adopted and the Pauli-exclusion principle is enforced through ideas of polymer excluded-volume. The atoms hydrogen through neon are examined without some approximations which permit cancellation of errors. Correlations are neglected in the interest of simplicity and comparisons are made with Hartree-Fock theory. The implications of the Pauli-exclusion potential and its approximate form are discussed, and the Pauli model is analyzed using scaling theory for the uniform electron density case where the correct form of the Thomas-Fermi quantum kinetic energy and the Dirac exchange correction are recovered.
\end{abstract}

\maketitle

\section{Introduction}

Density functional theory (DFT) is one of the most widely used and successful methods
for calculating structure and properties of many-body quantum systems. In DFT, a one-
particle density is the central quantity instead of a many-particle wave function, making
DFT computations orders-of-magnitude more tractable than wave function approaches. In
particular, Kohn-Sham DFT (KS-DFT), which uses orbital functions as a route to find the
density, can achieve chemical accuracy in many cases. 

It has been recently shown that polymer self-consistent field theory (SCFT) can also be used to study quantum many-body systems \cite{Thompson2019, Thompson2020, Sillaste2022, Thompson2022}. Instead of orbitals, SCFT uses propagators which are solutions to modified diffusion equations. The SCFT route has several
advantages: the propagators are real-valued functions in contrast to the complex-valued
orbitals used in KS-DFT, the diffusion equations are initial-value parabolic equations in
contrast to the elliptical boundary-value Kohn-Sham equation, the SCFT algorithm can be
made parallel in a straightforward way since the propagators do not span entire systems as
do KS orbitals, and the classical partition function derivation of the SCFT equations has
implications for the foundations of quantum mechanics \cite{Thompson2022}.

In previous work, SCFT has been used to study basic atomic \cite{Thompson2019, Thompson2020} and molecular \cite{Sillaste2022} systems using, for simplicity, orthogonal basis sets. In KS-DFT, non-orthogonal basis sets are often used, specifically, Gaussian basis functions. Gaussians have many advantages \cite{Helgaker1995}, including analytical solutions to molecular integrals and the ability to span the space with far fewer functions. Since the computational burden of SCFT, like KS-DFT, asymptotically scales with the cube of the number of basis functions, any reduction in the number of basis functions is very significant in terms of both computational runtimes and the numerical accuracy of results. 

In this work we consider the simple case of neutral atoms in order to develop non-orthogonal Gaussian basis sets for application to SCFT. We also implement improvements to the various approximations previously used in SCFT, specifically, we introduce an exact electron self-interaction correction. Existing methods of DFT have been developed over decades by hundreds of researchers in thousands of papers, and we cannot hope for SCFT to be immediately competitive with these mature methods in just a few publications \cite{Thompson2019, Thompson2020, Sillaste2022, Thompson2022}. Working towards this goal however, a necessary and non-trivial prerequisite is to demonstrate the efficacy of Gaussian basis functions within the SCFT approach. Although we will in some cases use approximations which give results that are less accurate or predictive than KS-DFT, these approximations are merely placeholders in order to implement
Gaussians in a simple way with minimal confounding factors. Our results are accurate by the standards of orbital-free DFT (OF-DFT) which is a more appropriate comparison since
we do not use orbitals in SCFT. The use of Gaussians in OF-DFT is not a mature field \cite{Chan_Cohen_Handy2001.article,Ryley_Withnall2021.article}, and studying atoms and molecules with OF-DFT can be very challenging \cite{Carter2012}. 

In Section \ref{Theory}, we summarize the main points of the SCFT approach and enumerate the approximations. We also give the spectral expansion of the SCFT governing equations in
the language of non-orthogonal basis sets which we then apply using Gaussian functions. In sections \ref{Results}, we give the binding energies for the atoms hydrogen through krypton calculated using Gaussian basis functions with three main approximations: a spherical averaging approximation, a shell-occupancy approximation and an approximate Pauli potential.
Radial electron density plots are given, and both the densities and binding energies are compared to Hartree-Fock (HF) theory. We further examine hydrogen through neon using
exact self-interaction expressions without the shell approximation in order to understand some instances of cancellation of error. A discussion of the implications of the various
approximations is given in Section \ref{Discussion}, with conclusions summarized in Section \ref{Conclusions}.

\section{Theory} \label{Theory}

\subsection{Summary of model and approximations}
The theory of polymer SCFT applied to quantum DFT has been described in detail elsewhere \cite{Thompson2019, Thompson2020, Sillaste2022}, and so here we merely summarize the governing equations and concentrate on describing the formalism in terms of non-orthogonal basis functions.

The free energy of a system of quantum particles, here electrons, subject to an external potential, such as the nucleus of an atom, can be derived from quantum statistical mechanics \cite{Thompson2019} following Kirkwood \cite {Kirkwood1933} and McQuarrie \cite{McQuarrie2000}, or equivalently, directly from a path integral formulation describing the conformations of a ring ``polymer'' in classical statistical mechanics with an imaginary time dimension representing the thermal energy $\beta = 1/k_BT$ of the system. This equivalency, called the quantum-classical isomorphism by Chandler and Wolynes \cite{Chandler1981}, was first outlined by Feynman \cite{Feynman1953b} and is the basis of several quantum simulation methodologies such as path integral Monte Carlo \cite{Ceperley1995}, centroid molecular dynamics \cite{Roy1999b,Zeng2014} and ring polymer molecular dynamics \cite{Habershon2013}. According to this picture, each quantum particle is a one dimensional thread extending into the thermal (imaginary time) dimension -- a thermal world-line. Each thread or trajectory follows a random walk in the absence of external influences. In polymer physics, this is called the Gaussian-thread model, since coarse-grained models of polymers in the melt obey the same random walk Gaussian statistics \cite{Matsen2002}. It is in this sense that each quantum particle is called a ``polymer'', or more exactly, a ring polymer, since Feynman showed that the initial spatial position of the thermal trajectory must be the same as the final position\footnote{In Feynman's application to bosons, the final positions may be permutations of the initial positions, but this can be handled instead through the Pauli potential \cite{Thompson2020}.} \cite{Feynman1953b}, which is the same mathematical condition that ring polymers obey. Note that real polymers are contours embedded in $\mathbb{R}^3$ whereas quantum particles are contours along the $\beta$ axis, which is the Wick rotation of imaginary time in Feynman's path integral action for quantum mechanics \cite{Feynman1953b}.

In this work, we will consider the simple case of electrons in neutral atoms, therefore the potentials the particles will be subject to include the Coulomb attraction of the nucleus,
the Coulomb electron-electron interaction, and the Pauli exclusion principle since electrons are Fermions. In the spirit of OF-DFT, the exclusion principle can be described by a
Pauli potential term. In the static 4-dimensional thermal-world-line picture, electrons are extended but classical objects, and so we use a classical model for the Pauli potential which
should become equivalent to the quantum exclusion principle, including exchange effects, when projected into three dimensions. Formally, it consists of a delta function repulsive
pseudo-potential acting between infinitesimal pieces of different ring polymer contours with equal values of imaginary time. To account for spin, this potential is felt only between
pairs of threads rather than between all threads, and so we group electrons in pairs so as to minimize the free energy of the system. This picture has been described in more detail
in reference \cite{Thompson2020}. In practice, calculations are easier if one averages over all imaginary time slices, so that repulsion is felt between all portions of polymers of different pairs rather than just between equivalent imaginary time slices. With this approximation to the Pauli model, the Pauli potential is written as \cite{Thompson2020}
\begin{equation}
U_P[\{n\}] = \frac{1}{2} \sum_{ij} \left(1-\delta_{ij}\right)g^{-1}_0 \int n_i({\bf r},\beta) n_j({\bf r},\beta) d{\bf r}   \label{UP}
\end{equation}
where $g_0^{-1}$ is a parameter with units of an inverse density of states that sets the strength of the Pauli potential. In (\ref{UP}), $n_{i}({\bf r})$ is the inhomogeneous electron density of pair $i$ and $\{n\}$ represents the set of all pair densities in the system. The double sum is over all pairs in the system, where a ``pair'' can have either 1 or 2 electrons in it. Note that (\ref{UP}) is an approximation to our Pauli potential model in that it applies a pseudo-potential Dirac delta repulsion between any segments of electron contours of different pairs rather than just between those segments at the same value of $\beta$. Thus, we should expect it to overestimate the Pauli repulsion, and we will see that the approximate form of (\ref{UP}) is a major limiting factor to the quantitative accuracy of the SCFT calculations. We improve the results by incorporating another approximation in which some limited shell information of atoms is incorporated. Specifically, we use a shell approximation where groups of more than two electrons are permitted according to known shell structure but ignoring subshells, and the Pauli repulsion acts only between electrons in different shells. More details of this shell approximation are given in Section \ref{Results}. The shell approximation has
serious consequences for the predictive capacity of our approach, but we note that shell structure is spontaneously predicted even without this approximation, as will be shown in
section \ref{Results}. The strength of the repulsive factor $g_0^{-1}$ should be a universal constant, but since (\ref{UP}) is an approximation, $g_0^{-1}$ will be close to, but not exactly, constant in applications. In this work, we use a fixed value of $g_0^{-1}=10$ (atomic units) which worked well for the atoms hydrogen through krypton using the approximate Pauli potential with the additional shell approximation. We use a third approximation by treating all atoms as spherically symmetric. From the work of Chowdhury and Perdew \cite{Perdew2021}, we do not expect this spherical averaging approximation to be significant. Other minor approximations, almost universally used in density functional theory calculations on lighter atoms, include the Born-Oppenheimer approximation, neglect of relativistic effects, and treating the nucleus as a point. We will not refer to these more minor assumptions further in this article.

The electron-electron interaction can be written in terms of pairs with a Fermi-Amaldi-type pre-factor \cite{Ayers2005} as 
\begin{equation}
U_{\rm ee}[\{n\}] = \frac{1}{2} \sum_{ij} \left(1-\frac{\delta_{ij}}{N_i} \right) \int \int n_i({\bf r},\beta) \mathcal{V}(|{\bf r} - {\bf r}^\prime|) n_j({\bf r}^\prime,\beta) d{\bf r} d{\bf r}^\prime   \label{Uee}
\end{equation}
where $\mathcal{V}(r)$ is the $1/r$ Coulomb potential and $N_i$ is the number of electrons in the $i$th pair. Note that as long as each pair has only one or two electrons in it, (\ref{Uee}) exactly accounts for self-interactions. If we use the shell approximation that allows more than two electrons per group, then self-interactions will no longer be exactly subtracted, and (\ref{Uee}) becomes more like a standard Fermi-Amaldi term and is a fourth approximation.

The final potential term for an atomic system is the Coulomb potential due to the nucleus -- the external potential. It depends only on the total electron density 
\begin{equation}
n({\bf r},\beta) = \sum_i n_i({\bf r},\beta)   \label{totn}
\end{equation}
rather than on the pairs individually. It is
\begin{equation}
U_{\rm ext}[n] = -N \int \int n({\bf r},\beta) \mathcal{V}(|{\bf r} - {\bf r}^\prime|) \rho_{\rm ion}({\bf r}^\prime) d{\bf r} d{\bf r}^\prime   \label{Uext}
\end{equation}
where $N$ is the atomic number of a neutral atom. We treat the nucleus as a point particle with $\rho_{\rm ion}({\bf r}) = \delta({\bf r})$, so that (\ref{Uext}) becomes 
\begin{equation}
U_{\rm ext}[n] = -N \int n({\bf r},\beta) \mathcal{V}(|{\bf r}|) d{\bf r}  .  \label{Uext2}
\end{equation}

From reference \onlinecite{Thompson2020}, the canonical ensemble free energy for this model is
\begin{equation}
F[\{n\},\{w\}] = -\frac{1}{\beta}\sum_i N_i \ln Q_i(\beta) - \sum_i \int d{\bf r} w_i({\bf r},\beta) n_i({\bf r},\beta)  + U[\{n\}]     .\label{FE1}
\end{equation}
where
\begin{equation}
U[\{n\}] = U_{\rm ext}[n]+U_{\rm ee}[\{n\}]+U_P[\{n\}]  \label{Utot}
\end{equation}
is the total potential and $w_i({\bf r},\beta)$ are fields conjugate to each density pair $n_i({\bf r},\beta)$. $Q_i(\beta)$ are single particle partition functions for each electron pair and are functionals of the fields $w_i({\bf r},\beta)$. Expressions for the fields and the densities can be found by varying (\ref{FE1}) with respect to all the field and density functions. Details on this procedure can be found in references \onlinecite{Matsen2020,Matsen2006,Qiu2006,Fredrickson2002,Fredrickson2006,Schmid1998}. The results are
\begin{eqnarray}
n_i({\bf r},\beta) &=& \frac{N_i}{Q_i} q_i({\bf r},{\bf r},\beta)  \label{dens1}  \\
w^P_i({\bf r},\beta) &=& g_0^{-1} \sum_{j \ne i} n_j({\bf r},\beta) \label{wP} \\
w^{\rm ee}_i ({\bf r},\beta) &=& \int \left[ n({\bf r}^\prime,\beta)-\frac{n_i({\bf r}^\prime,\beta)}{N_i}\right]\mathcal{V}(|{\bf r} - {\bf r}^\prime|) d{\bf r}^\prime  \label{wee}   \\
w^{\rm ext} ({\bf r}) &=& -N \int \mathcal{V}(|{\bf r} - {\bf r}^\prime|) \rho_{\rm ion}({\bf r}^\prime) d{\bf r}^\prime \label{wext}
\end{eqnarray}
where 
\begin{equation}
Q_i(\beta) = \int q_i({\bf r},{\bf r},\beta) d{\bf r}   \label{Q1}
\end{equation}
and 
\begin{equation}
\frac{\partial q_i({\bf r}_0,{\bf r},s)}{\partial s} = \frac{\hbar^2}{2m} \nabla^2 q_i({\bf r}_0,{\bf r},s) - w_i({\bf r},\beta) q_i({\bf r}_0,{\bf r},s)  \label{diff1}
\end{equation}
with
\begin{equation}
w_i({\bf r}) = w^P_i({\bf r},\beta) + w^{\rm ee}_i ({\bf r},\beta) + w^{\rm ext} ({\bf r})    \label{wtot}
\end{equation}
subject to the initial conditions
\begin{equation}
q_i({\bf r}_0,{\bf r},0) = \delta({\bf r}-{\bf r}_0)   .  \label{init1}
\end{equation}
The single particle partition functions (\ref{Q1}), and the initial condition (\ref{init1}), are defined slightly differently from reference \onlinecite{Thompson2020} in that a factor of volume $V$ is not included. We do this in order to let $V \rightarrow \infty$, which removes issues of finite size effects from our calculations. The polymeric character of quantum particles is captured in the real-valued modified diffusion equation (\ref{diff1}) which gives the unnormalized probability of finding a particle at position ${\bf r}$ for inverse temperature $\beta$ if it is known to be at position ${\bf r}_0$ for high temperatures ($\beta = 0$).

\subsection{Spectral Representation}

The main burden in solving the self-consistent set of equations (\ref{dens1})-(\ref{init1}) are the diffusion equations (\ref{diff1}). Real space methods are not practical due to the double spatial dependence on ${\bf r}$ and ${\bf r}_0$, but with a bilinear spectral expansion, (\ref{diff1}) is computationally no more demanding to solve than a diffusion equation of a single spatial variable \cite{Thompson2019}. Therefore we expand all spatially dependent functions in terms of a basis set $f_k ({\bf r})$, $k=1, 2, 3, \cdots$ according to 
\begin{equation}
g({\bf r}) = \sum_k g_k f_k ({\bf r})  \label{exp1}
\end{equation}
where $g({\bf r})$ is an arbitrary spatially dependent function and $g_k$ are the components of that function with respect to the basis set $f_k ({\bf r})$. For functions of two spatial variables, like equation (\ref{diff1}), we use a bilinear expansion
\begin{equation}
g({\bf r}_0, {\bf r}) = \sum_k g_{kl} f_k ({\bf r}_0)  f_l ({\bf r})    .  \label{exp2}
\end{equation}
An infinite number of terms are required in expansions (\ref{exp1}) and (\ref{exp2}) to represent the functions precisely, but we truncate the series at a finite number sufficient to produce a chosen numerical accuracy. In references \onlinecite{Thompson2019, Thompson2020, Sillaste2022}, we used orthogonal basis functions with the symmetry of the systems. For atoms however, this required us to use a very large number of basis functions. An industry standard is the use of Gaussian basis sets for atoms and molecules, but these sets are non-orthogonal which complicates the spectral representation of the SCFT equations. It is the main purpose of this paper to give the SCFT equations in terms of non-orthogonal basis functions and show that Gaussian functions are highly effective in the solution of the SCFT representation of localized quantum systems such as atoms.

Using (\ref{exp1}) and (\ref{exp2}) to write the SCFT equations in terms of non-orthogonal basis functions produces several molecular integrals. These are:
\begin{eqnarray}
S_{ij} &=& \int f_i({\bf r}) f_j({\bf r}) d{\bf r}  \label{overlap}  \\
L_{ij} &=& \int f_i({\bf r}) \nabla ^2 f_j({\bf r}) d{\bf r}  \label{Laplacian}  \\
\Gamma_{ijk} &=& \int f_i({\bf r}) f_j({\bf r}) f_k({\bf r}) d{\bf r}  \label{Gamma}
\end{eqnarray}
which we call the overlap matrix, the Laplacian matrix and the Gamma tensor. For orthogonal basis sets, these would simplify, for example, the overlap matrix would become a simple Kronecker delta. For non-orthogonal basis sets such as Gaussians however, this is not the case. Throughout this paper, in order to clarify the notation, we will reduce the use of indices by writing spectral equations using a matrix notation without indices and write only the electron pair index explicitly. Therefore, the integrals (\ref{overlap})-(\ref{Gamma}) are written in matrix notation as ${\bf S}$, ${\bf L}$ and ${\bf \Gamma}$.

Expanding equations (\ref{dens1})-(\ref{init1}) according to (\ref{exp1}) and (\ref{exp2}) gives\footnote{It is helpful to rephrase (\ref{wee}) and (\ref{wext}) in terms of the Poisson equation in order to expand them spectrally.}
\begin{eqnarray}
{\bf S} {\bf n}_i &=& \frac{N_i}{Q_i} {\bf \Gamma} {\bf q}_i\label{dens2}   \\
{\bf w}^P_i &=&  g_0^{-1} \left({\bf n} - {\bf n}_i \right)   \label{wP2}  \\
{\bf L} {\bf w}^{\rm ee}_i &=& -4\pi {\bf S}\left({\bf n} - \frac{{\bf n}_i}{N_i}\right)   \label{wee2}   \\
{\bf L}{\bf w}^{\rm ext} &=& 4\pi N {\bf S}{\bf \bm{\rho}}  \label{wext2}
\end{eqnarray}
where ${\bf n}_i$, ${\bf w}^P_i$ and ${\bf w}^{\rm ee}_i$ are the $i$th pair electron component vectors of the density, Pauli field and electron-electron field, respectively, and ${\bf n}$,  ${\bf w}^{\rm ext}$ and $\bm{\rho}$ are the component vectors of the total density, external field and the ion density, respectively. Additionally,
\begin{equation}
Q_i = {\rm Tr} ({\bf S}{\bf q_i})  \label{Q2}
\end{equation}
with the matrix ${\bf q}_i$ for the $i$th pair given by
\begin{equation}
\frac{d{\bf q}_i}{ds}  = {\bf S}^{-1} {\bf A}_i {\bf q}_i  \label{diff2}
\end{equation}
where 
\begin{equation}
{\bf A}_i = \frac{1}{2}{\bf L} - {\bf \Gamma}{\bf w}_i   \label{A1}
\end{equation}
and
\begin{equation}
{\bf w}_i = {\bf w}_i^P + {\bf w}_i^{\rm ee} + {\bf w}^{\rm ext}  .  \label{wtot2}
\end{equation}
In equations (\ref{dens2})-(\ref{wtot2}), we have suppressed $\beta$ dependencies for clarity. The non-orthogonal spectral representation of the initial condition (\ref{init1}) works out to be
\begin{equation}
{\bf q}(0) = {\bf S}^{-1} \,,  \label{init2}
\end{equation}
which allows us to write the solutions to eqn. (\ref{diff2}), with the aid of the expression for the identity matrix $\bf{I} = \bf{U}_i^{T}\bf{S}\bf{U}_i$, as
\begin{equation}
{\bf q}_i(s) = {\bf U}_i{\bf D}_i{\bf U}_i^T\,,  \label{diff3} 
\end{equation}
where ${\bf U}_i$ is the matrix of generalized eigenvectors corresponding to the matrix pencil (${\bf A}_i, {\bf S}$) for the $i$th pair of electrons and ${\bf D}_i$ is the diagonal matrix comprised of the vector of generalized eigenvalues $\bm{\lambda}_i$ given by
\begin{equation}
{\bf D}_i = {\rm e}^{\bm{\lambda}_i s} \,.  \label{D1}
\end{equation}
The transition to the generalized eigenvalue problem addresses a subtle issue that arises when working with non-orthogonal basis sets: the two matrices $\bf{S}$ and $\bf{A}_i$ are both separately Hermitian, but the matrix product $\bf{S}^{-1}\bf{A}_i$ is in general not Hermitian, forcing the computer to abandon the Hermiticity of the problem. Therefore, any computer implementation of the model ought to incorporate the generalized eigenvalue problem if the most accurate results are to be harvested. 

Equation (\ref{diff3}) allows us to rewrite (\ref{Q2}) in the more computationally convenient form 
\begin{equation}
Q_i = {\rm Tr} ({\bf D}_i) \,,  \label{Q3}
\end{equation}
by using the cyclic property of the matrix trace and the expression for the identity matrix $\bf{I} = \bf{U}_i^{T}\bf{S}\bf{U}_i$. The equations (\ref{dens2})-(\ref{Q3}) are then solved for the electron density components ${\bf n}_i$ for a given external potential by choosing a basis set with the symmetry of the system under study and using it to specify the molecular integrals (\ref{overlap})-(\ref{Gamma}) and the external potential (\ref{wext2}). The algorithm for the self-consistent solution for the density component vectors ${\bf n}_i$ is given below, as is the choice of the basis set used in this work.

\subsection{Basis functions and numerics}

For atoms and molecules, Gaussian basis functions are a standard choice due to their ability to represent localized electron densities -- a review of the advantages and disadvantages of Gaussians can be found in reference \onlinecite{Helgaker1995}. Importantly, the ability to perform molecular integrals such as (\ref{overlap})-(\ref{Gamma}) analytically means that computations can be very efficient. The atom-centred normalized basis function set used in this work is given by
\begin{equation}  
f_i(r) = \left(\frac{2\alpha_i}{\pi}\right)^\frac{3}{4} {\rm e}^{-\alpha_i r^2}  \label{Gauss}
\end{equation}
where $i = 1, 2, 3 \cdots$ is an integer index (not the electron pair index), $\alpha_i$ is the (inverse) width of the $i$th basis function and $r$ is the radial distance from the nucleus. We let the basis functions depend only on $r$ since Chowdhury and Perdew suggest the angular dependence of electron densities for neutral atoms is not significant \cite{Perdew2021}. Thus we are using a spherical averaging approximation in this work. In terms of (\ref{Gauss}), the molecular integrals (\ref{overlap})-(\ref{Gamma}), when written in component form, become
\begin{eqnarray}
S_{ij} &=& \left[\frac{4 \alpha_i\alpha_j}{(\alpha_i+\alpha_j)^2}\right]^\frac{3}{4}  \label{overlap2}  \\
L_{ij} &=& -\frac{6S_{ij}\alpha_i\alpha_j}{(\alpha_i+\alpha_j)}   \label{Laplacian2}  \\
\Gamma_{ijk} &=& \left[\frac{8\alpha_i\alpha_j\alpha_k}{\pi(\alpha_i+\alpha_j+\alpha_k)^2}\right]^\frac{3}{4}  \label{Gamma2} 
\end{eqnarray}
which, aside from the specification of the widths $\bm{\alpha}$, completes the information needed to solve the spectral equations (\ref{dens2})-(\ref{wext2}) self-consistently. In particular, the external potential components (\ref{wext2}) become
\begin{equation}
{\bf L}{\bf w}^{\rm ext} = 4\pi N \left(\frac{2 \bm{\alpha}}{\pi}\right)^\frac{3}{4}  .  \label{wext3}
\end{equation} 
Since Gaussian functions form an over-complete basis set, the widths $\bm{\alpha}$ need to be chosen carefully. This can be done in a number of ways, often involving contracted basis sets \cite{Huzinaga1996}. In this work however, since we are using simple atomic systems to demonstrate the efficacy of Gaussian basis functions for SCFT, we will avoid more sophisticated methods of choosing $\bm{\alpha}$, and instead use an even-tempered basis set \cite{Schmidt1979, Cherkes2009}. Our even-tempered scheme differs from the canonical scheme outlined in \onlinecite{Schmidt1979}, since the exponents are determined by evenly partitioning the logarithmic space between a minimum and maximum exponent of our choosing. This approach is motivated from the procedure used in \onlinecite{Cherkes2009} and can be justified using the encouraging results from references \onlinecite{Cherkes2009} and \onlinecite{Schmidt1979} showing the relative linearity of optimized basis exponent sets. The choice of the minimum and maximum exponents is informed by how accurate the total electron and pair electron numbers are compared to their known values, and changes depending on how many basis functions are included in the set. It is possible to introduce linear dependence problems by defining the basis set in too narrow an interval, and the minimum and maximum widths should also technically depend on the number of electrons in the system since the inhomogeneities in the electron density due to shell structure could alter the optimal range of exponents or increase the deviation of the sequence from linearity. However, we kept the minimum and maximum widths fixed for our calculations and our results did not show any signs of significant issues due to this factor.

\begin{figure}
\centering
\includegraphics[height=8cm, width=13cm]{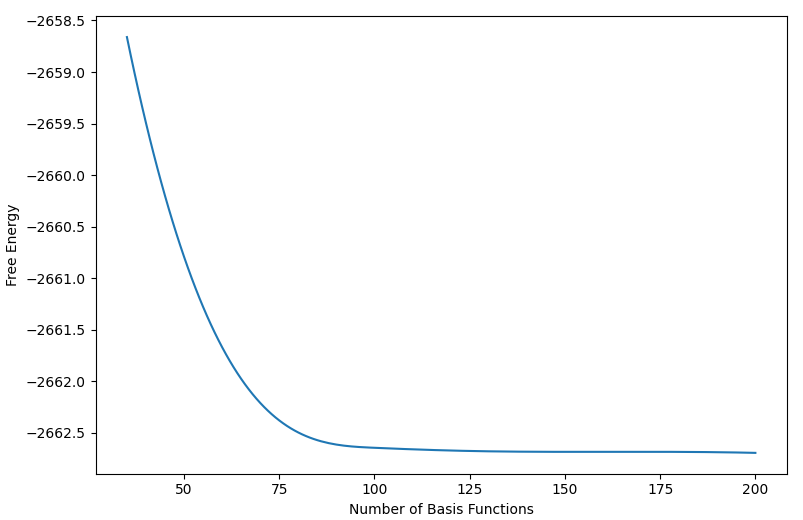}
\caption[Krypton Free Energy Convergence with Basis Set Size]{The convergence of the free energy for krypton with the number of Gaussian basis functions used. The plot clearly demonstrates the exponential convergence of the Gaussians mentioned in the text. Note that basis function numbers between 50 and 175 have been interpolated using a cubic spline interpolator for expediency, and values outside this range have been extrapolated.} \label{fig:basis}
\end{figure}

For each of our calculations, we used $M$ basis functions $\bm{\alpha}$ with $\alpha_1 = 10^{-15}$, $\alpha_{M} = 10^{11}$ and $M-2$ more Gaussians with widths logarithmically spaced (even-tempered) between the maximum and minimum values according to
\begin{equation}
\log_{10}{\alpha_i} = \log_{10}{\alpha_{i-1}} + \Delta   \label{spacing}
\end{equation}
where
\begin{equation}
\Delta = \frac{\log_{10}{\alpha_{\rm max}} - \log_{10}{\alpha_{\rm min}}}{M-1}  \,.  \label{Delta}
\end{equation}
For each atom, we did runs with $M=\{50, 75, 100, 125, 150, 175\}$ in order to estimate the basis set truncation error on the atomic binding energy, as shown in figure \ref{fig:basis} for krypton.
All runs were performed with $\beta \ge 100$, which we found to approach the ground state $\beta \rightarrow \infty$ solution to within acceptable numerical errors. We checked this on various runs by using also $\beta \ge 200$ which produced no consequential numerical change in the results. The self-consistent calculation was continued until a specific $L^2$ norm between the input and output field components, weighted by the density, agreed to less than one part in $10^{-7}$. Certain runs were tested to a self-consistent tolerance of less than one part in $10^{-10}$. No consequential numerical change resulted. Of the three sources of numerical error -- basis set truncation, finite value of $\beta$ for ground state calculations and self-consistent tolerance -- the basis set truncation was the limiting factor on the reported results. We did not use extrapolation techniques, such as Richardson extrapolation, for our binding energy estimates since the model approximations limited our accuracy compared to HF much more than numerical factors.

The self-consistent iterations were carried out by expanding the external potential in terms of the Gaussian basis set to give the components (\ref{wext2}), and initial guesses for electron densities were also expanded to give density component vectors ${\bf n}_i$. The fields (\ref{wP2}) and (\ref{wee2}) were calculated and summed with (\ref{wext2}) to form an input guess for the field components. The matrices ${\bf A}_i$ were constructed using (\ref{A1}) and (\ref{wtot2}) and the generalized eigenvalue problems involving ${\bf S}^{-1}$ and ${\bf A}_i$ were then solved in order to find the propagator matrices from (\ref{diff3}) and (\ref{D1}). The single particle partition functions (\ref{Q3}) were also found, which allowed the electron density components to be evaluated from (\ref{dens2}), and from these, output field components from (\ref{wP2}), (\ref{wee2}) and (\ref{wext2}). The input and output field components were compared, and if they agreed within an acceptable numerical tolerance, the self-consistent calculation was terminated. Otherwise, the input and output field components were combined to form a new input guess and the cycle was repeated. More details of the convergence criteria used are given in appendix \ref{L2}.

\section{Results}   \label{Results}

\begin{figure}
\begin{tabular}{ccc}
SCFT ($g_0^{-1} = 3.0$) & SCFT ($g_0^{-1} = 10.0$) & Hartree-Fock \\
(a) & (b) & (c) \\
\includegraphics[width=0.35\textwidth]{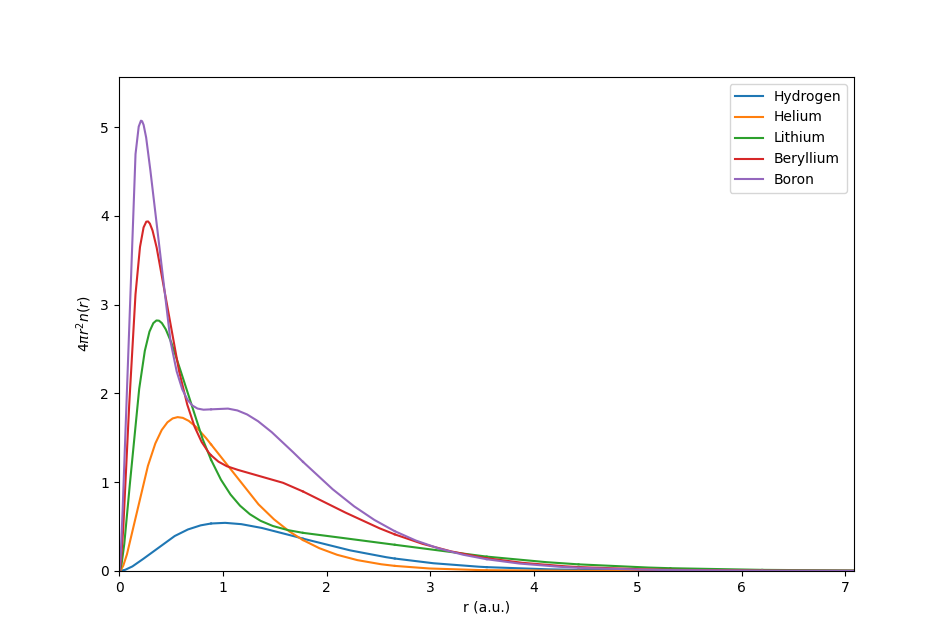} & \includegraphics[width=0.35\textwidth]{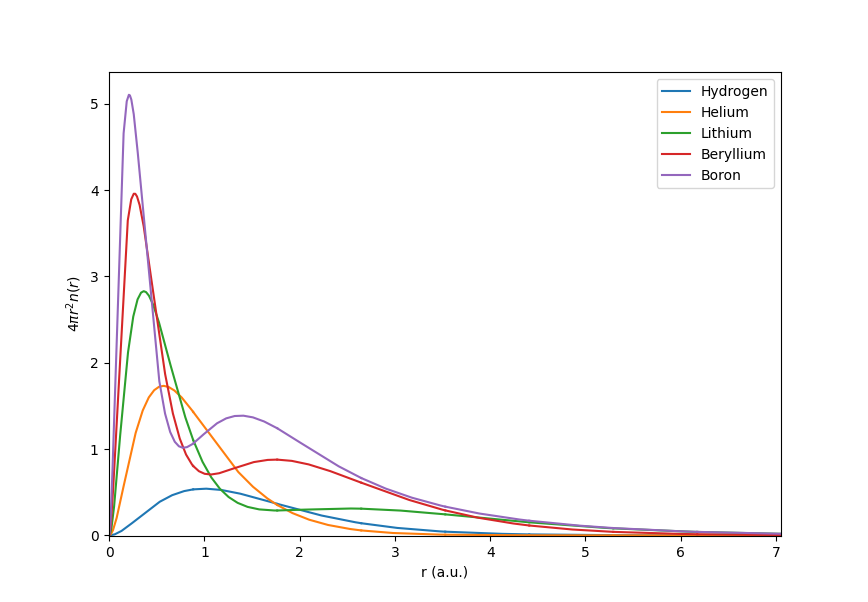} & \includegraphics[width=0.3\textwidth]{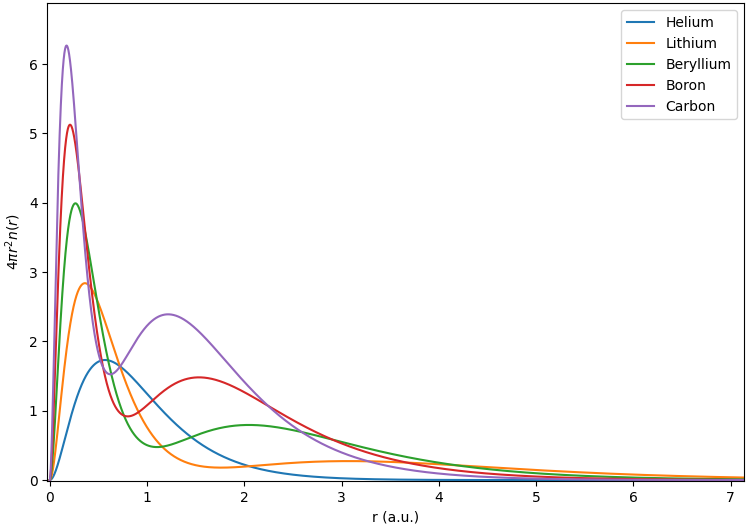} \\
(d) & (e) & (f)\\
\includegraphics[width=0.35\textwidth]{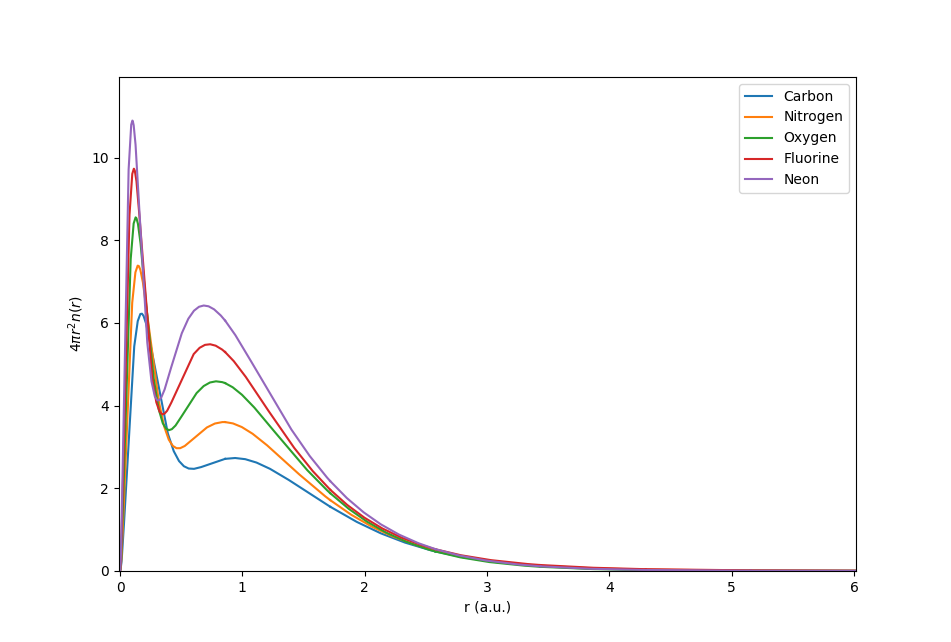} & \includegraphics[width=0.35\textwidth]{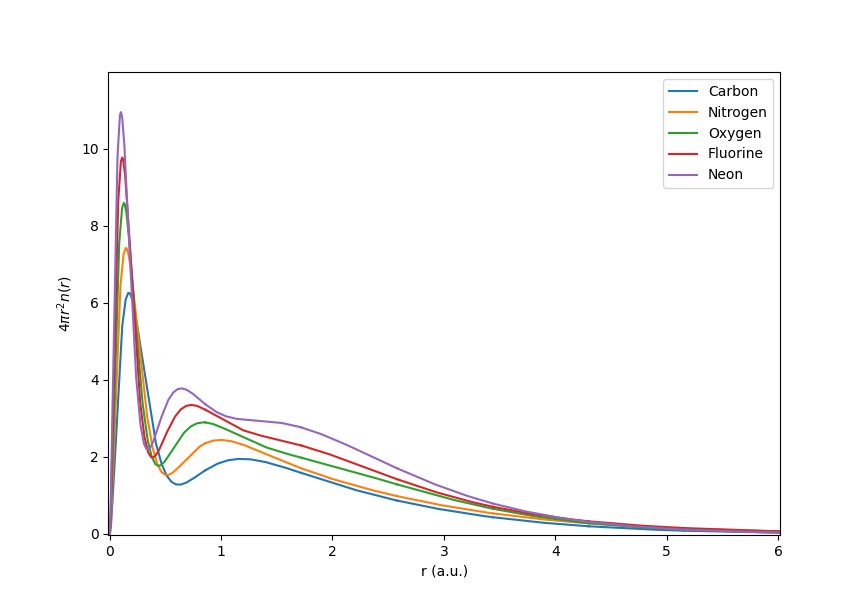} & \includegraphics[width=0.32\textwidth]{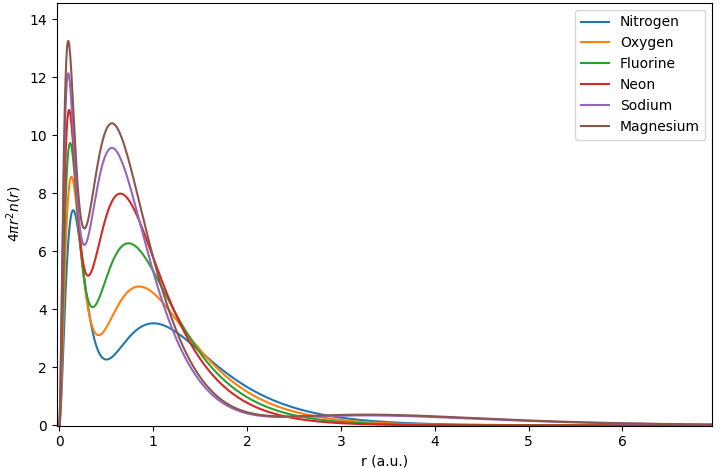} \\
\end{tabular}
\caption{Plots of radial electron densities for SCFT without the shell approximation using $g_0^{-1} = 3.0$ (a), (d); $g_0^{-1} = 10.0$ (b), (e); and HF \cite{Strand1964} (c), (f). Legends denote the elements.}
\label{fig:H-Ne}
\end{figure}

Tables \ref{tab:energies1} and \ref{tab:energies2} gives atomic binding energies for a selection of elements using $g_0^{-1} = 10$, each table using different approximations. Atomic units are used throughout. The decimal places on the SCFT free energies reflect the numerical errors on the calculations, which were limited by the finite basis set. The SCFT energies can be compared to Hartree-Fock (HF) values since, for simplicity, we are not including any correlation effects in this work. Therefore, HF can be taken as ``exact'' in the absence of correlations. Table \ref{tab:energies1} uses two main approximations in addition to the neglect of correlations: we assume the electron density is spherically symmetric (spherical averaging approximation) and we use the approximate expression of the Pauli potential given by equation (\ref{UP}) that averages over imaginary time slices in calculations of the exclusion repulsion. 
\begin{longtable}{p{0.05\textwidth} p{0.08\textwidth} p{0.27 \textwidth} p{0.21\textwidth} p{0.18\textwidth} p{0.15\textwidth} }
\caption{SCFT atomic binding energies with spherical averaging approximation and Pauli approximation compared to Hartree-Fock values. Estimated SCFT numerical uncertainty is restricted to the last digit of each entry.} \label{tab:energies1} \\
\hline
$N$ & Atom & Electronic configuration & SCFT Free Energy & Hartree-Fock \cite{Koga1996} & {\% Difference}  \\ 
\hline
\endfirsthead
\caption{(Continued.)} \\
\hline
$N$ & Atom & Electronic configuration & SCFT Free Energy & Hartree-Fock \cite{Koga1996} & {\% Difference}  \\ 
\hline
\endhead
\bottomrule
\hline
\endfoot
\hline
\endlastfoot
1 & H & 1s(1) & 0.49999998 & 0.500000000 & $4.0 \times 10^{-6}$ \\
2 & He & 1s(2)& 2.8616800 & 2.861679996 &  $1.4 \times 10^{-7}$ \\ 
3 & Li & [He]2s(1) & 7.468419 & 7.432726931 & 0.48 \\
4 & Be & [He]2s(2) & 14.702194 & 14.57302317 & 0.89 \\ 
5 & B & [He]2s(2)2p(1) & 24.66953 & 24.52906073 & 0.57 \\
6 & C & [He]2s(2)2p(2) & 37.56774 & 37.68861896 & 0.32 \\ 
7 & N & [He]2s(2)2p(3) & 53.4071 & 54.40093421 & 1.83 \\ 
8 & O & [He]2s(2)2p(4) & 72.3335 & 74.80939847 & 3.31 \\ 
9 & F & [He]2s(2)2p(5) & 94.3264 & 99.40934939 & 5.11 \\ 
10 & Ne & [He]2s(2)2p(6) & 119.5084 & 128.5470981 & 7.03 
\end{longtable}
In table \ref{tab:energies1}, SCFT energies for elements hydrogen and helium can be seen to exactly agree with HF within our numerical accuracy. This makes sense since both atoms are known to be spherically symmetric and neither is affected by the Pauli potential. Also, whereas in our previous work \cite{Thompson2020, Sillaste2022} we used the Fermi-Amaldi approximation to correct for self-interactions in the electron-electron Coulomb expression \cite{Ayers2005}, we can now notice that since we are treating the electrons in pairs, the Fermi-Amaldi correction applied to each pair individually becomes exact. Therefore, since lithium and beryllium are also spherically symmetric, the deviations from HF of 0.5\% and 0.9\%, respectively, can be completely attributed to the approximate Pauli expression -- there are no other major approximations left except the neglect of correlations, which are also neglected in HF theory. Elements boron through neon should be additionally affected by the spherical averaging approximation, but we assume this will be a very small effect following the findings of reference \onlinecite{Perdew2021}. For lighter elements, the radial electron densities agree very well with HF and spontaneously produce shell structure as shown in figures \ref{fig:H-Ne} (a), (b), (d), and (e).The main contribution to the growing deviation between SCFT and HF with increasing atomic number is then the Pauli potential expression, and this will be examined in more detail in the Discussion section.\footnote{We do not populate the pairs according to Hund's rule because the pair propagators are not in general the same as orbitals. In any case, the Pauli approximation is too coarse to allow us to distinguish between energies of various $p$ shell populations.} For now, we can see that the energies for elements nitrogen through neon become poor compared to the lighter elements, and this trend will continue for elements beyond neon. Also, the electron densities will cease to be even qualitatively correct for heavier elements. Therefore, a correction is needed to improve the Pauli potential.

\begin{figure}
\begin{tabular}{cc}
SCFT & Hartree-Fock \\
(a) & (b) \\
\includegraphics[width=0.4\textwidth]{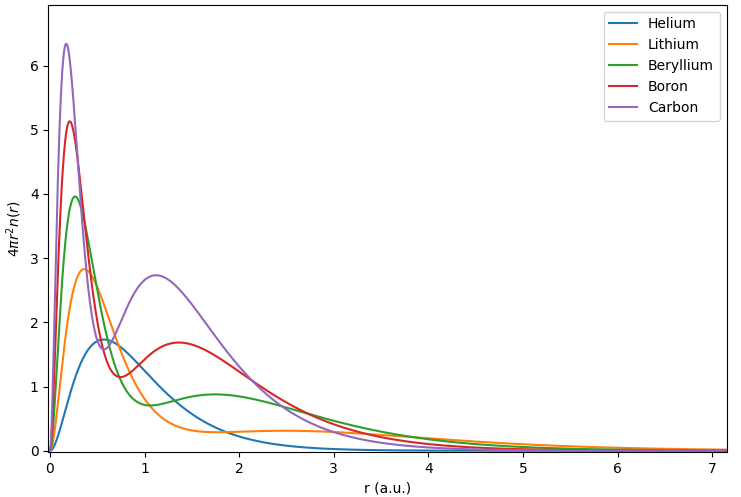} & \includegraphics[width=0.4\textwidth]{./figures/HF_He-C.png} \\
(c) & (d) \\
\includegraphics[width=0.4\textwidth]{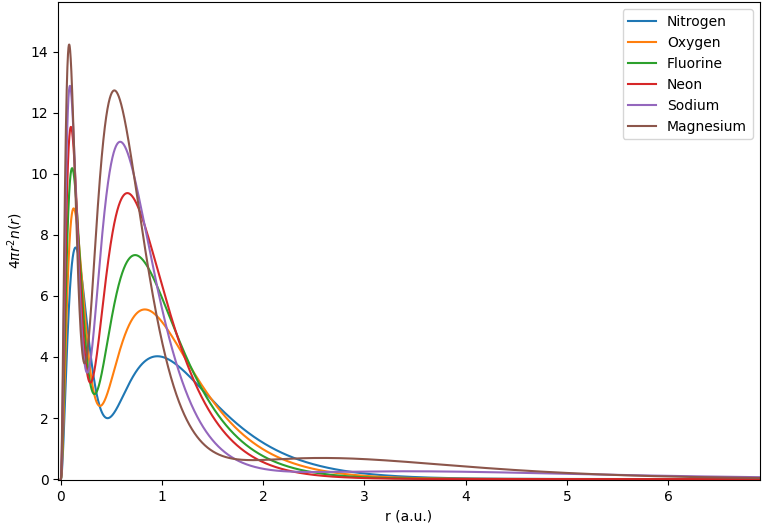} & \includegraphics[width=0.4\textwidth]{./figures/HF_N-Mg.png} \\
(e) & (f) \\
\includegraphics[width=0.4\textwidth]{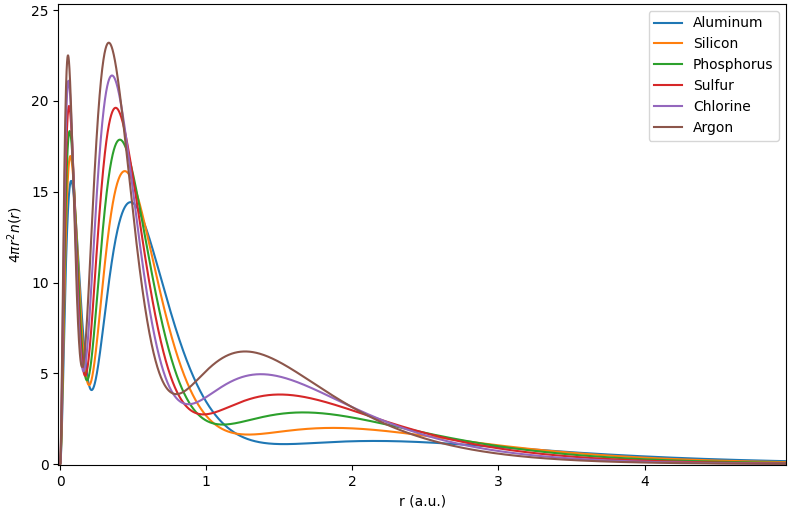} & \includegraphics[width=0.4\textwidth]{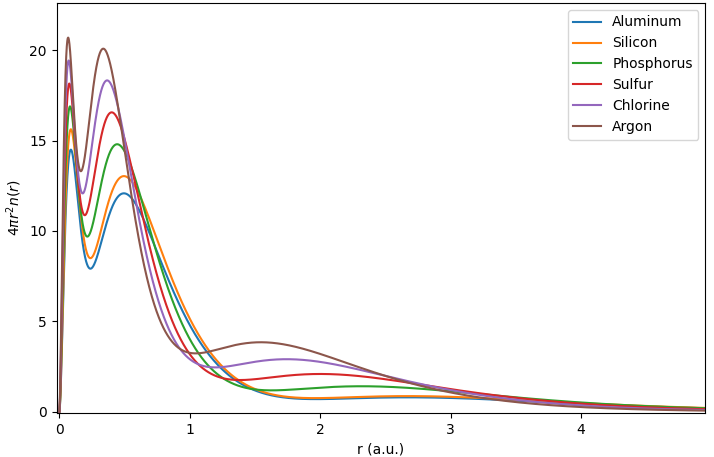} 
\end{tabular}
\caption{Plots of SCFT (a), (c), (e) and HF \cite{Strand1964} (b), (d), (f) radial electron densities. Legends denote the elements.}
\label{fig:He_Ar}
\end{figure}

\begin{figure}
\begin{tabular}{cc}
SCFT & Hartree-Fock \\
(a) & (b) \\
\includegraphics[width=0.4\textwidth]{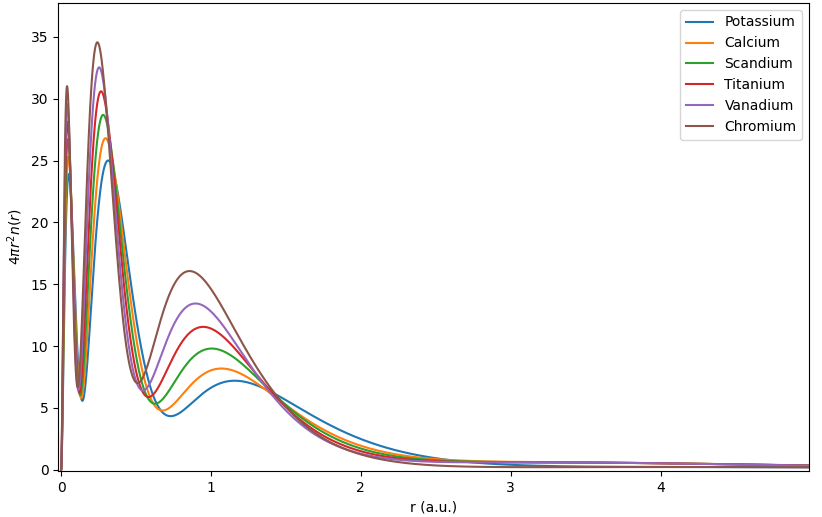} & \includegraphics[width=0.4\textwidth]{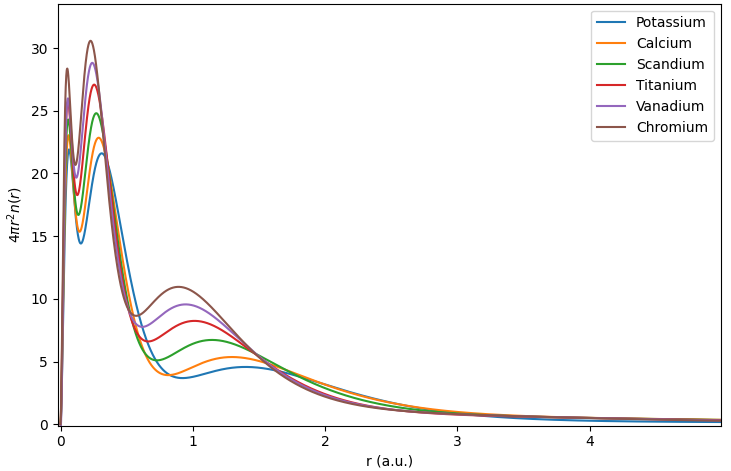} \\
(c) & (d) \\
\includegraphics[width=0.4\textwidth]{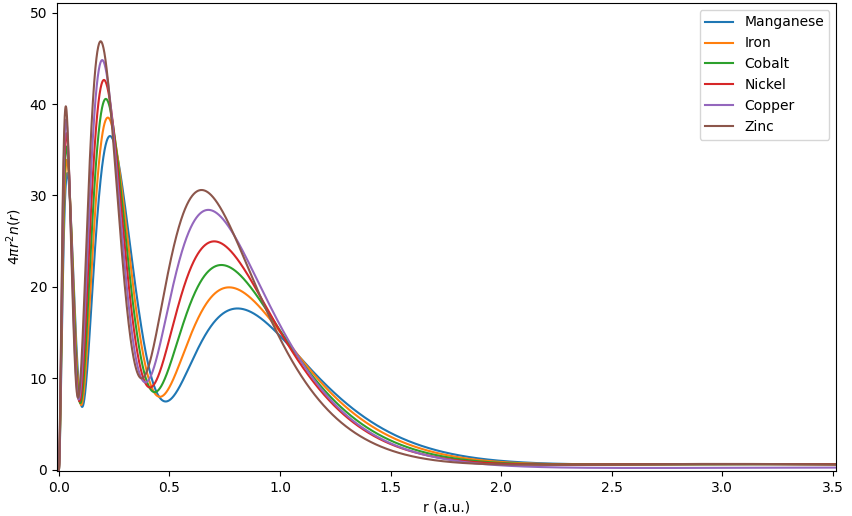} & \includegraphics[width=0.4\textwidth]{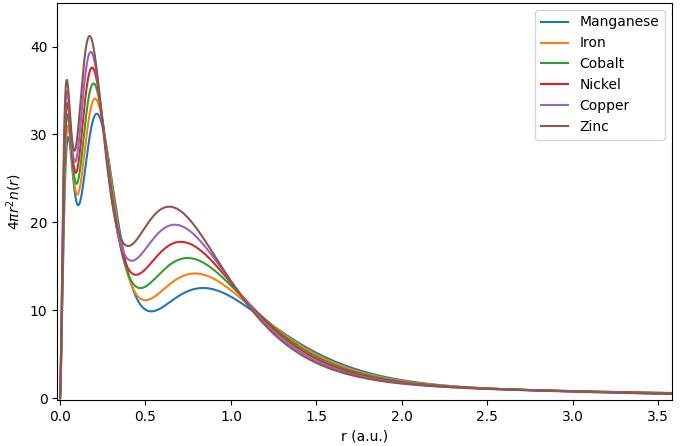} \\
(e) & (f) \\
\includegraphics[width=0.4\textwidth]{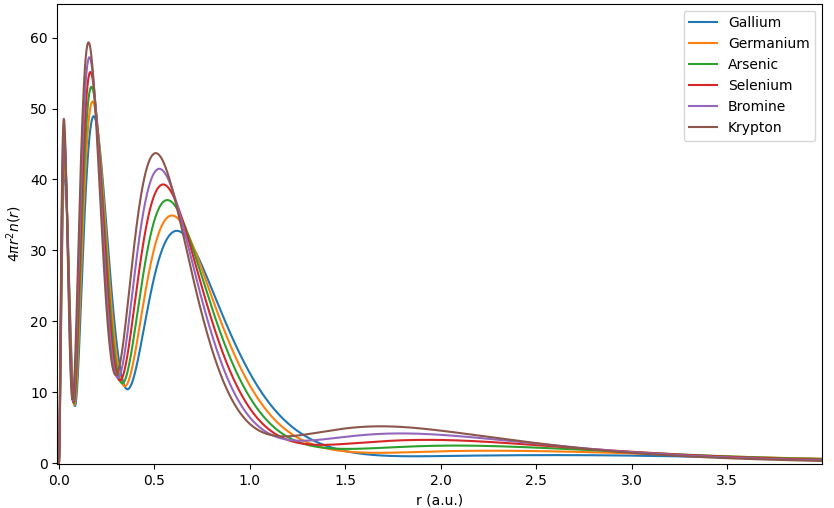} & \includegraphics[width=0.4\textwidth]{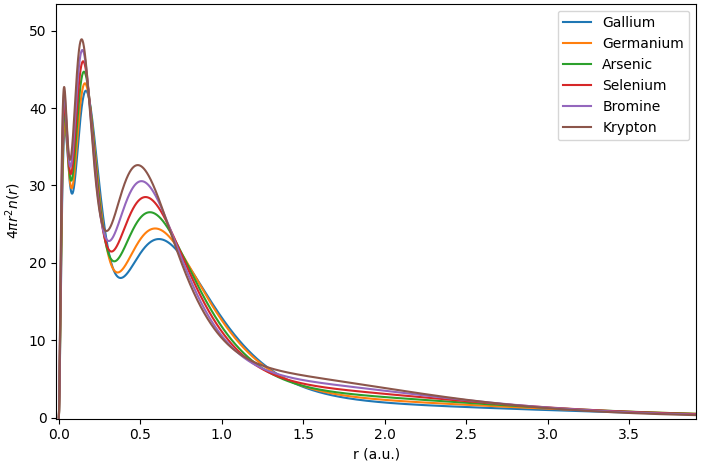}
\end{tabular}
\caption{Plots of SCFT (a), (c), (e) and HF \cite{Strand1964} (b), (d), (f) radial electron densities. Legends denote the elements.}
\label{fig:K_Kr}
\end{figure}

Table \ref{tab:energies2} gives the atomic binding energies for the first 36 elements of the periodic table, and includes the same two main approximations as table \ref{tab:energies1}, plus a third assumption. Following our previous work \cite{Thompson2020}, we ignore sub-shell structure of atoms and consider the Pauli potential to apply only between electrons in different shells.
\begin{longtable}{p{0.05\textwidth} p{0.08\textwidth} p{0.27 \textwidth} p{0.21\textwidth} p{0.18\textwidth} p{0.15\textwidth} }
\caption{Atomic binding energies with spherical averaging approximation, Pauli approximation and shell approximation compared to Hartree-Fock values. Estimated SCFT numerical uncertainty is restricted to the last digit of each entry.} \label{tab:energies2} \\
\hline
$N$ & Atom & Electronic configuration & SCFT Free Energy & Hartree-Fock \cite{Koga1996} & {\% Difference}  \\ 
\hline
\endfirsthead
\caption{(Continued.)} \\
\hline
$N$ & Atom & Electronic configuration & SCFT Free Energy & Hartree-Fock \cite{Koga1996} & {\% Difference}  \\ 
\hline
\endhead
\bottomrule
\hline
\endfoot
\hline
\endlastfoot
1 & H & 1s(1) & 0.49999998 & 0.500000000 & $4.0 \times 10^{-6}$ \\
2 & He & 1s(2)& 2.8616800 & 2.861679996 &  $1.4 \times 10^{-7}$ \\ 
3 & Li & [He]2s(1) & 7.46842 & 7.432726931 & 0.48 \\
4 & Be & [He]2s(2) & 14.70219 & 14.57302317 & 0.89 \\ 
5 & B & [He]2s(3) & 24.90400 & 24.52906073 & 1.53 \\
6 & C & [He]2s(4) & 38.40323 & 37.68861896 & 1.90 \\ 
7 & N & [He]2s(5) & 55.52627 & 54.40093421 & 2.07 \\ 
8 & O & [He]2s(6) & 76.59989 & 74.80939847 & 2.39 \\ 
9 & F & [He]2s(7) & 101.95295 & 99.40934939 & 2.56 \\ 
10 & Ne & [He]2s(8) & 131.91735 & 128.5470981 & 2.62 \\ 
11 & Na & [Ne]3s(1) & 165.5246 & 161.8589116 & 2.26 \\
12 & Mg & [Ne]3s(2) & 203.3441 & 199.6146364 & 1.87 \\
13 & Al & [Ne]3s(3) & 245.5076 & 241.8767073 & 1.50 \\
14 & Si & [Ne]3s(4) & 292.1387 & 288.8543625 & 1.14 \\ 
15 & P & [Ne]3s(5) & 343.3593 & 340.7187810 & 0.77 \\ 
16 & S & [Ne]3s(6) & 399.2911 & 397.5048959 & 0.45 \\ 
17 & Cl & [Ne]3s(7) & 460.0566 & 459.4820724 & 0.13 \\ 
18 & Ar & [Ne]3s(8) & 525.7794 & 526.8175128 & 0.20 \\ 
19 & K & [Ar]4s(1) & 595.9661 & 599.1647868 & 0.53 \\
20 & Ca & [Ar]4s(2) & 670.9221 & 676.7581859 & 0.86 \\
21 & Sc & [Ar]3s(1)4s(2) & 752.0400 & 759.7357180 & 1.01 \\
22 & Ti & [Ar]3s(2)4s(2) & 838.6134 & 848.4059970 & 1.15 \\
23 & V & [Ar]3s(3)4s(2) & 930.7714 & 942.8843377 & 1.28 \\
24 & Cr & [Ar]3s(5)4s(1) & 1030.1603 & 1043.356376 & 1.26 \\
25 & Mn & [Ar]3s(5)4s(2) & 1132.365 & 1149.866252 & 1.50 \\
26 & Fe &  [Ar]3s(6)4s(2) & 1242.066 & 1262.443665 & 1.61 \\
27 & Co & [Ar]3s(7)4s(2) & 1357.883 & 1381.414553 & 1.70 \\
28 & Ni & [Ar]3s(8)4s(2) & 1479.953 & 1506.870908 & 1.79 \\
29 & Cu & [Ar]3s(10)4s(1) & 1610.739 & 1638.963742 & 1.72 \\
30 & Zn & [Ar]3s(10)4s(2) & 1743.398 & 1777.848116 & 1.94 \\
31 & Ga & [Ar]3s(10)4s(3) & 1881.819 & 1923.26010 & 2.15 \\
32 & Ge & [Ar]3s(10)4s(4) & 2026.074 & 2075.359734 & 2.37 \\
33 & As & [Ar]3s(10)4s(5) & 2176.225 & 2234.238654 & 2.60 \\
34 & Se & [Ar]3s(10)4s(6) & 2332.335 & 2399.867612 & 2.81 \\
35 & Br & [Ar]3s(10)4s(7) & 2494.468 & 2572.441333 & 3.03 \\
36 & Kr & [Ar]3s(10)4s(8) & 2662.684 & 2752.054977 & 3.25 
\end{longtable}
Thus, for example, we would treat neon as having an electronic structure of [He]2s(8) instead of [He]2s(2)2p(6), with the Pauli potential being felt only between the two 1s and eight 2s electrons, but not between electrons in the same shell. Since, as will be shown in the discussion, we know that our Pauli expression (\ref{UP}) is too strong, the shell approximation will go some way to compensate. It requires aspects of known shell structure of the atoms as input, so it severely compromises the predictive power for the current application, and we use
it as a stopgap to develop and demonstrate the efficacy of Gaussian basis functions in SCFT. The shell approximation could be replaced in the future by a more sophisticated model of
the Pauli potential --- some options are mentioned in Section \ref{Conclusions} --- but for situations that don't require chemical accuracy, as in many OF-DFT calculations, the assumption of very basic atomic shell structure information could remain a pragmatic engineering input to many molecular calculations. The shell approximation will introduce error on the self-interaction correction, since we are no longer exclusively dealing with singlet or pair electrons in each shell. The compensation for large over-estimations of the Pauli potential in the heavier elements is worth this trade-off. Elements H through Be are unchanged from table \ref{tab:energies1}, but for elements B through Kr, the shell approximation improves our results. Percent deviations from HF increases for B through Ne, but remain below 3\%. For Na through Cl, the percent deviation steadily decreases as the binding energies become less and less over-bound. Errors are below 2.5\% for all these elements and below 1\% for many of them. For Ar, the error is only 0.2\%, but we can see some quantitative issues with the electron density where the minima are deeper than expected -- electron density plots for all the elements of table \ref{tab:energies2} compared with HF electron densities are given in figures \ref{fig:He_Ar} and \ref{fig:K_Kr}. 
There is some cancellation of errors in the binding energies, which is not surprising given the Pauli and shell approximations. Ar is slightly under-bound, and all elements throughout the rest of the periodic table from this point become close to monotonically more under-bound. Percent deviations from HF remain below 1\% for elements P through Ca, and below 2\% for Sc through Zn. Only Br and Kr have deviations from HF above 3\%, but for the heavier elements, the radial electron densities can be seen to deviate more from HF. The binding energies and electron densities of all elements of the periodic table beyond Kr have been computed but are not reported since the quantitative under-binding continues to grow.

\section{Discussion} \label{Discussion}

The deviation between SCFT and HF binding energies is not dominated by numerical errors, but overwhelmingly by approximations to the SCFT model. If we assume that the spherical averaging approximation has only a small effect, we are left with the Pauli expression (\ref{UP}) as responsible for almost all deviation from HF. The strength of the Pauli repulsion is characterized by $g_0^{-1}$, and in principle, this should be a universal quantity -- any single experiment would set this value. This is true however inasmuch as we have an exact mathematical form for the Pauli potential. As discussed in reference \cite{Thompson2020}, expression (\ref{UP}) does not act only between equivalent imaginary time slices, but is rather an average interaction over all slices. It will thus be too strong a repulsion. In references \onlinecite{Thompson2020, Thompson2022} we used a value of $g_0^{-1} \approx 5.74$ based on a comparison with the homogeneous electron gas Thomas-Fermi expression, but we expected this to be valid only within an order-of-magnitude due to the neglect of density dependence just mentioned \cite{Thompson2020}. In this work, given our approximate expression (\ref{UP}), we treated $g_0^{-1}$ as a phenomenological parameter and found a value of $g_0^{-1} = 10$ gave good results for elements H through Kr as shown in table \ref{tab:energies2}. The value 10 was initially chosen because it gave excellent results for argon, which lies directly in the middle of the range of elements considered in this work. The value for $g_0^{-1}$ was chosen to be a constant rather than a function of the electron number $N$ in order to minimize fine-tuning of parameters in the model. One could conceivably fit a function to the HF data for each $N$ and arbitrarily approach the exact binding energy predicted by HF theory, but this still would not produce the correct density profiles because the structure of the approximate Pauli-exclusion field differs by more than just a constant factor from the exact field. Treating $g_0^{-1}$ phenomenologically parallels the practice in polymer physics where the Flory-Huggins parameter $\chi$ has a well defined meaning in terms of the change of energy of a polymer melt upon exchange of monomers, but is usually treated as a fitting parameter in practice \cite{Matsen2020}. 

One may question whether it is just the approximate form of the mathematical expression (\ref{UP}) for the Pauli potential that is at fault, or if the underlying model of treating the exclusion principle as excluded-volume between polymer threads in the higher dimensional thermal-space is responsible for the quantitative disagreement between SCFT and HF. Following reference \onlinecite{Thompson2020}, one can use scaling theory arguments of de Gennes \cite{deGennes1979} to show that both the Thomas-Fermi quantum kinetic energy and Dirac exchange are obtained from the higher dimensional classical model in the case of high uniform density quantum systems. This is shown in appendix \ref{TFD}, and is evidence in favour of the classical thread model for the Pauli-exclusion principle. In addition to the scaling behaviour, systematic investigations of the exact Pauli potential $U_P$ and its associated field $w^P(\bm{r}, \beta)$, undertaken by Levy and Ou-Yang \cite{Levy_Ou-Yang1988.article}, have produced a number of constraints that must be satisfied by the exact Pauli-exclusion field or any approximation thereof. The relevant constraints are as follows:
\begin{eqnarray}
&w^P(\bm{r}, \beta) \geq 0\ \ , \ \ \Lim{|\bm{r}|\rightarrow \infty} w^P(\bm{r}, \beta) = 0\ \ , \ \ \int \mathrm{d}\bm{r}\, w^P(\bm{r}, \beta)n(\bm{r}, \beta) < \infty\ \ , \nonumber \\[1.5ex]  &\ \ w^P(\bm{r}, \beta) = 0 \ \ \text{for} \ \ N=2\ \ , \ \ w^P[\lambda^3n](\lambda\bm{r}, \beta) = \lambda^2w^P[n](\lambda\bm{r}, \beta)\,, \label{pauli_cons}
\end{eqnarray}
where in the last criterion $\lambda$ is a scale factor and the functional dependence of the field on the density has been explicitly reinstated. We are thus provided with some useful benchmarks to examine the severity of approximating the Pauli potential eqn. (\ref{UP}) in this work. The first constraint is satisfied by eqn. (\ref{wP}), since the total electron density and the individual pair electron densities obey the same constraint, meaning the total electron density must be larger than each pair density by virtue of eqn. (\ref{totn}). The second constraint is also satisfied by eqn. (\ref{wP}), since the electron density goes to zero far away from the nucleus. The satisfaction of the first two constraints combined with the properties of the electron density imply that the third constraint is satisfied by eqn. (\ref{wP}) as well. The fourth constraint is trivially satisfied because each pair can be comprised of at most 2 electrons and eqn. (\ref{wP}) is zero for only 1 pair. By inspection, the last constraint clearly does not hold for eqn. (\ref{wP}), suggesting that the approximation used in this work does not quite scale correctly with coordinate scaling. This last point is discussed in reference \onlinecite{Thompson2020}, where it becomes clearer that the approximate Pauli potential used in this model overestimates the excluded-volume interactions by precisely the amount required to fulfill the last constraint in eqn. (\ref{pauli_cons}). The accuracy of our current results are obviously worse than KS-DFT, but this can be remedied by improving the Pauli potential approximation. There are other potential calculations that can be performed to improve the numerical representation of this model within SCFT, and we hope to investigate these in possible future work --- see Section \ref{Conclusions}.

We can compare the results of table \ref{tab:energies1} to our previous work of reference \onlinecite{Thompson2020} where we used orthogonal spherical Bessel functions in a finite spherical box to calculate the binding energies of H through Ar. The current results for light elements H through O are much better than in reference \onlinecite{Thompson2020} due to better basis function resolution\footnote{In reference \onlinecite{Thompson2020}, we compared to NIST rather than HF, but the small correlations included in the NIST values don't change our conclusions.}. In particular residual finite size effects seem to have affected He in reference \onlinecite{Thompson2020} whereas in this present work, the Gaussian basis functions allow us to use an infinite radius and, as mentioned, our He binding energy result agrees with HF to within less than a few ten-millionths of a percent. Percent errors start to become comparable to reference \onlinecite{Thompson2020} at about F and Ne. Results improve significantly for Na through Ar compared to reference \onlinecite{Thompson2020} with deviations dropping monotonically until Ar, compared to reference \onlinecite{Thompson2020} where deviations increase over this range. Some of this is due to the change in our choice of the value of $g_0^{-1}$ and some due to the removal of residual finite size effects and greater basis function resolution. We see that the use of non-orthogonal Gaussians is superior in almost every way for localized systems like atoms compared to orthogonal basis sets, consistent with current understanding from other DFT approaches. This indicates that most algorithmic improvements of general DFT methods can likely be ported to the polymer-thread model. 

Given the success of Gaussian basis functions for the polymeric method of solving DFT, it is relevant to consider whether the SCFT methodology may hold advantages for some applications compared to KS or OF-DFT. These established methods are much more mature, and SCFT is not yet competitive with KS-DFT in terms of accuracy; implementing Gaussian basis functions is a necessary early step towards that goal. If the shell approximation can be removed in favour of a more quantitatively accurate Pauli potential, then both the accuracy and predictive power of SCFT could improve significantly. The shell approximation is a blunt instrument which limits the predictive power of SCFT. Shell structure can be observed even without the shell approximation however, as shown in figures \ref{fig:H-Ne} (a), (b), (d), and (e), so if the placeholder Pauli expression (\ref{UP}) can be replaced with an improved potential, SCFT
could compete with or exceed KS-DFT in terms of accuracy and predictive power. Some possibilities for this are mentioned in section \ref{Conclusions}.

There are other more immediate advantages of SCFT compared to KS-DFT. The diffusion equations (\ref{diff1}) are initial-value parabolic partial differential equations, which are generally easier to solve numerically than the boundary value elliptic Kohn-Sham equation \cite{Kohn1965}. In particular, the propagators of(\ref{diff1}) are real-valued, compared to the orbitals of KS theory which are complex functions. Also, when solving the propagators for pairs of localized electrons, the functions are also localized in the vicinity of a given electron pair as opposed to orbitals which, in general, span the whole system. For large molecules, this means the propagators can be solved using only basis functions localized near the pairs, following
Kohn's \"near-sighted\" principleThis means that for large systems the SCFT method scales linearly with the number of electrons. Linear scaling can also be achieved with orbitals, but not directly. The SCFT method is a particularly facile way of achieving linear scaling,although we do not give a numerical example in this work --- this is a potential future direction
involving calculations of large molecules. Still along these lines, each of the propagators can be calculated independent of the others, so the SCFT approach is \"embarrassingly parallel\".
Again, we have not implemented this feature here since we are studying only single atoms with a focus on demonstrating Gaussian basis sets. If accuracy is not of primary concern, a coarser implementation of the Pauli potential can reduce the computation to a single diffusion equation, in which case the SCFT method becomes similar to existing OF-DFT methods and would be computationally very much faster than KS-DFT.

Examples of Gaussian basis functions being used in orbital-free DFT (OF-DFT) are rare --– see for example references \onlinecite{Chan_Cohen_Handy2001.article} and \onlinecite{Ryley_Withnall2021.article}. In these cases, the quality of predictions of binding energies for atoms vary compared to our results, although we uniquely find exact agreement with HF for hydrogen and helium, that is, for systems where a Pauli potential is not needed. The OF Gaussian methods also do not show shell structure, unlike in our case. Were we to replace our pair-based Pauli potential with one more similar to those used in other work \cite{Chan_Cohen_Handy2001.article,Ryley_Withnall2021.article}, then we would need only a single diffusion equation and our method would scale in the same way as OF-DFT but could maintain exact results for hydrogen and helium. Thus we get the same computational advantages as OF-DFT in that limiting case, but with exact HF results for the lightest atoms. Unlike OF-DFT based on an Euler-Lagrange equation, we can get shell structure if we use multiple diffusion equations. Alternatively, we could divide systems up into core and valence electrons, and use an intermediate number of diffusion equations for a compromise between accuracy and speed. We have already done something similar in our study of diatomic molecules \cite{Sillaste2022}.

From a practical perspective, the SCFT algorithm is relatively easy to implement. We needed only about 200 lines of code, including comments, to produce the results presented
here. An experienced researcher could conceivably implement an SCFT program in less time than it takes to learn to use some commercial or academic packages. This gives more flexibility and control through direct manipulation of the source code.

\section{Conclusions and Future Work}  \label{Conclusions}

Non-orthogonal Gaussian basis functions are extremely effective for representing polymer SCFT when applied to atomic quantum many-body systems. Using between 100 and 200 basis functions, greater numerical accuracy is achieved than using well over 1000 orthogonal basis functions such as spherical Bessel functions \cite{Thompson2020}. Since the asymptotic computational burden scales as the cube of the number of basis functions, the Gaussian solution algorithm is well over 200 times faster. In fact, the Gaussian approach is faster still due to an increased robustness in the algorithm and analytical values for the overlap matrix, Laplacian matrix and the Gamma tensor. Since Gaussian basis functions extend to infinity, the computation does not refer to a computational volume which eliminates residual finite-size effects. These results are consistent with the advantages of Gaussian basis functions in orbital-based DFT \cite{Helgaker1995, Huzinaga1996, Schmidt1979, Cherkes2009}, and so it is reasonable to assume further improvements could be realized by incorporating other known DFT techniques such as contracted basis sets in which several Gaussians are combined in fixed linear combinations that are then used as basis functions. This would be effective in reducing the number of basis function for molecular systems. The SCFT study of molecular systems will require angular dependence in the basis functions, and Cartesian Gaussians would be practical for this. Molecular computations would then follow the basic prescription of other DFT algorithms and we have already calculated the necessary molecular integrals, implemented 3D Cartesian Gaussian basis functions with multiple centres, and validated the mapping of contracted basis sets to known atomic electron densities. Full molecular implementation will be the subject of future work. Angular dependence of basis functions could also be implemented in the atomic case using spherical harmonics in order to validate Chowdhury and Perdew's conclusions about the role of spherical symmetry
breaking in neutral atoms. This will also be the subject of future investigations. 

Other potential future work includes the effects of temperature dependence and improvements to the Pauli potential. The SCFT methodology automatically includes temperature dependence and issues of thermal decoherence can readily be studied. The Pauli potential was represented in this work by an average over all imaginary time slices of the polymer contour, but there are ways to avoid this approximation. One option is to divide each thread contour into sections and allow interactions only between the sections | this would follow numerical techniques used in multi-block copolymers \cite{Matsen1994}. Another approach would be to include exchange in a more traditional way by allowing conformations that interchange final positions between different threads, as mentioned by Feynman in his original description \cite{Feynman1953b}.As mentioned, the coarse Pauli approximation (\ref{UP}) is a placeholder suitable for semi-quantitative calculations. Given that self-interactions can be represented exactly in the SCFT formalism, if the Pauli potential could be improved there would be no approximations left except the neglect of correlations. These too can be included in future work so that chemical accuracy is not an unreasonable goal. For larger molecules however, the ease and speed of the average Pauli potential approximation used in this work would probably be more practical. In the future, it would not be complicated to parallelize the SCFT algorithm since it is naturally suited to it. 

\section*{Acknowledgements}

This research was financially supported by the Natural Sciences and Engineering Research Council of Canada (NSERC). 

\appendix
\section{Convergence Criterion}  \label{L2}
Simple Picard iteration was used to combine input and output field components with a mixing parameter in the range of 0.1-0.5, supplemented with Anderson mixing using up to 20 histories when the convergence slowed to below $10\%$ change between iterations -- see reference \onlinecite{Thompson2012} for more details on Anderson mixing and the convergence strategy that we used. Since field values are unimportant in regions where the density approaches zero, we stabilized our convergence by weighting the deviation between the input and output field components with a factor of the density components. This procedure can be seen as a type of inner product between arbitrary functions $g({\bf r})$ and $h({\bf r})$ weighted by a function $n({\bf r})$ defined as
\begin{equation}
\left(g({\bf r}),h({\bf r})\right) = \int n({\bf r}) g({\bf r}) h({\bf r}) d{\bf r} \,,  \label{inner}
\end{equation}
where the arbitrary functions are replaced by the field deviation $|w^{\rm output}_i(\bm{r}, \beta)-w^{\rm input}_i(\bm{r}, \beta)|$ and the weight by the electron density $n({\bf r}, \beta)$ in this work. Equation (\ref{inner}) with field deviations substituted takes on the form, after spectral expansion:
\begin{equation}
\sum_i\text{Tr}\left[{\bf n}_i \left({\bf w}_i^{\rm output} - {\bf w}_i^{\rm input}\right)\left({\bf w}_i^{\rm output} - {\bf w}_i^{\rm input}\right){\bf \Gamma}\right] \equiv \sum_{i}\text{Tr}\left({\bf n}_i{\bf d}_i{\bf d}_i{\bf \Gamma}\right) \,. \label{dev1}
\end{equation} 
The total deviation $d_{\rm tot}$ used in this work is then given by the square root of eqn. (\ref{dev1}) divided by a normalizing factor 
\begin{equation}
\mathcal{N} \equiv \sum_{i} \text{Tr}\left({\bf n}_i {\bf w}_i^{\rm output}{\bf w}_i^{\rm output}{\bf \Gamma}\right) \,, \label{dev2}
\end{equation}
which we express as
\begin{equation}
d_{\rm tot} =  \sqrt{\left( \frac{\sum_{i}\text{Tr}\left({\bf n}_i{\bf d}_i{\bf d}_i{\bf \Gamma}\right)}{\mathcal{N}} \right)}\,,  \label{dtot1}
\end{equation}
where the sum runs over all pairs (shells).

\section{Quantum Kinetic Energy and Exchange for a Uniform Quantum System}   \label{TFD}

The Pauli potential model of polymer threads interacting through excluded volume in the thermal dimension can be tested in the high density uniform limit using scaling theory. This allows one to remove the approximation averaging over the contour variable and consider repulsions only between equivalent imaginary time slices. We can show that the model gives the correct scaling behaviour for the Thomas-Fermi (TF) quantum kinetic energy and Dirac exchange. The TF result has already been discussed in the appendix of reference \onlinecite{Thompson2020}, but for completeness, we repeat and expand the derivation here.

For high electron number density, each electron polymer contour will encounter many other contours and will be repulsed by all of them. Thus we can treat the electron contour as confined within a hyper-cylinder due to the other threads in the 4D thermal-space. We are ignoring spin, which will only add an overall factor of 2 and so will not change the scaling behaviour. We imagine dividing the thread up into $Z$ very small segments, each with an average contour length of $a$. An unconfined Gaussian thread will have a size $R_0 = Z^\frac{1}{2} a$.\footnote{Muller et al. \cite{Muller1996} find that the influence of the topological constraints (ring polymer architecture as opposed to a linear polymer) on the static properties of isolated rings appears not to alter the size exponents.} Following de Gennes \cite{deGennes1979}, the confinement of the thread will change its conformational entropy (which is equivalent to the quantum kinetic energy in the 3D picture \cite{Thompson2019}) by an amount $\Delta S$ which is to leading order proportional to $Z$.\footnote{See for example reference \onlinecite{Doi2001} page 19.} de Gennes also notes that $\Delta S$ should be dimensionless and depend only on the ratio of the size of the thread $R_0$ to the diameter of the cylinder $D$. Thus 
\begin{equation}
\Delta S = - \left(\frac{R_0}{D}\right)^y  \sim Z^\frac{y}{2}  . \label{ent1}
\end{equation}
Since $\Delta S$ should be linear in $Z$, we must have $y=2$. The Pauli energy per confined thread is then
\begin{equation}
\frac{U_P}{N} \sim D^{-2}  . \label{UP2}
\end{equation}
where $N$ is the number of threads in the system. The diameter $D$ of the hyper-cylinder in 4D is the diameter of a sphere in 3D, since each slice of the hypercylinder is just 3D space, so $D \sim n_0^{-1/3}$, where $n_0 = N/V$ and $V$ is the volume of the system. Thus (\ref{UP2}) becomes
\begin{equation}
\frac{U_P}{V} \sim n_0^{5/3}   \label{UP3}
\end{equation}
in agreement with the Thomas-Fermi quantum kinetic energy expression.

The TF expression accounts only for the conformational entropy of a confined thread and not the energy of the confinement, that is, the energy of contact with the cylinder. The cylinder energy is given by the number of contacts between the thread and the cylinder wall, so again following de Gennes, we have
\begin{equation}
E \sim \delta f_b Z  \label{Dirac1}
\end{equation}
where $\delta $ is the strength of the interaction between the thread and the wall and $f_b$ is the fraction of chain segments that touch the wall. This fraction will be $f_b \cong a/D$ since most of the segments spread over the diameter $D$ will not touch, only those within a segment length $a$. Therefore, the energy per thread will be 
\begin{equation}
\frac{E}{N} \sim \frac{1}{D} . \label{Dirac2}
\end{equation}
Again noting that the diameter $D$ of the 4D hyper-cylinder is the diameter of a 3D sphere $D \sim n_0^{-1/3}$, we get
\begin{equation}
\frac{E}{N} \sim n_0^{1/3}  . \label{Dirac3}
\end{equation}
The energy density (per volume) is then 
\begin{equation}
\frac{E}{V} \sim n_0^{4/3}  . \label{Dirac4}
\end{equation}
This is the expected Dirac exchange energy. The total Thomas-Fermi-Dirac expression is then 
\begin{equation}
\frac{E}{V} = A n_0^{5/3} -  B n_0^{4/3}   \label{TFD2}
\end{equation}
where $A$ and $B$ are constants.

\section*{Acknowledgements}
This research was financially supported by the Natural Sciences and Engineering Research Council of Canada (NSERC). 

\bibliography{DFTbibliography}

\end{document}